\documentclass[paper]{ieice}
\usepackage[dvipdfmx]{graphicx,xcolor}
\usepackage[fleqn]{amsmath}
\usepackage{multirow}
\usepackage[psamsfonts]{amssymb}
\usepackage{amsxtra}
\usepackage{amsmath}
\usepackage{ulem} 
\usepackage{soul}
\usepackage{threeparttable,comment}
\usepackage{cite,array}
\usepackage{subfig}
\usepackage{mathrsfs,color}
\usepackage{mathtools}

\usepackage{threeparttable,bm,amsmath,graphicx,amssymb,comment,array,multirow,threeparttable,cite,calligra,color}
\field{}
\title{Image Manipulation Specifications on Social Networking Services for
Encryption-then-Compression Systems} \authorlist{
\authorentry[chuman-tatsuya@ed.tmu.ac.jp]{Tatsuya CHUMAN}{n}{tmu}\MembershipNumber{}
\authorentry[iida-kenta1@ed.tmu.ac.jp]{Kenta IIDA}{s}{tmu}\MembershipNumber{}
\authorentry[warit-sirichotedumrong@ed.tmu.ac.jp]{Warit Sirichotedumrong}{n}{tmu}\MembershipNumber{}
\authorentry[kiya@tmu.ac.jp]{Hitoshi KIYA}{f}{tmu}\MembershipNumber{}}
\affiliate[tmu]{The authors are with Tokyo Metropolitan University, Hino-shi, 191-0065 Japan}
\received{2015}{1}{1}
\revised{2015}{1}{1}
\begin{document}
\maketitle
\begin{summary}
Encryption-then-Compression (EtC) systems have been proposed to securely
transmit images through an untrusted channel provider. In this study, EtC
systems were applied to social media like Twitter that carry out image
manipulations. The block scrambling-based encryption schemes used in EtC systems
were evaluated in terms of their robustness against image manipulation on social
media. The aim was to investigate how five social networking service (SNS)
providers, Facebook, Twitter, Google+, Tumblr and Flickr, manipulate images and to determine whether the encrypted images
uploaded to SNS providers can avoid being distorted by such manipulations. In an
experiment, encrypted and non-encrypted JPEG images were uploaded to various SNS
providers. The results show that EtC systems are applicable to the five SNS
providers.
\end{summary}
\begin{keywords}
JPEG, image encryption, EtC system, social media
\end{keywords}

\section{Introduction}
The use of images and video sequences has greatly increased because of rapid
growth of the Internet and multimedia systems. A lot of studies on secure,
efficient and flexible communications have been
reported\cite{huang2014survey,lagendijk2013encrypted,zhou2014designing}. For
securing multimedia data, full encryption with provable security (as with RSA
and AES) is the most secure option. However, many multimedia applications have
been seeking a trade-off in security to enable other requirements, e.g.,\, a
small amount of processing, bitstream compliance, or signal processing in
the encrypted domain. Here, perceptual encryption schemes have been studied
as ways of achieving this
trade-off\cite{Zeng_2003,Ito_2008,Kiya_2008,Ito_2009,Tang_2014}.
\par
Image encryption sometimes must be performed prior to image compression in
certain practical scenarios such as secure image transmission through an
untrusted channel provider. This framework is carried out by
Encryption-then-Compression (EtC)
systems\cite{zhou2014designing,Erkin_2007,nimbokar2014survey}. In this paper,
we focus on EtC systems, although the traditional way of securely
transmitting images is to use a Compression-then-Encryption (CtE) system.
However, most studies on EtC systems assume the use of their own compression
schemes that have no compatibility with international standards such as
JPEG\cite{Johnson_2004,Liu_2010,Zhang_2010,Hu_2014,zhou2014designing}. In this
paper, we focus on block scrambling-based image encryption schemes which have
compatibility with international compression
standards\cite{watanabe2015encryption,kurihara2015encryption,KURIHARA2015,KuriharaBMSB,Kuri_2017}.
\par
On the other hand, almost all social networking service (SNS) providers support
JPEG, one of the most widely used image compression
standards\cite{JPEG_1991}.
However, JPEG images are uploaded to SNS providers by users on the assumption that the  uploaded images in SNS servers are trustable, so the privacy of uploaded images is not under the control of the users. Nevertheless, there is no way to protect the uploaded images, because SNS providers generally manipulate them.
Although some papers have studied image manipulation on social
media\cite{Caldelli_2017,giudice2016_arxive,Moltisanti2015_ICIAP}, e.g.,
alternation of image filenames or headers of JPEG images, the recompression
parameters and the
conditions of image manipulation remain unclear.
Therefore, we investigated how SNS providers manipulate images and whether
EtC systems are applicable to their methods.
\par
We uploaded a lot of images to five SNS providers, Facebook,
Twitter, Google+, Tumblr and Flickr, to examine the robustness of EtC systems.
We found that encrypted images including some block distortion due to image manipulations on some SNS providers greatly reduce the quality
of the decrypted images. Otherwise, we confirmed that the EtC systems are
applicable to five SNS providers.

\section{Preparation}
\subsection{Necessity of EtC systems}

The importance of this work is to point out that most encryption schemes,
such as RSA and AES, are not applicable to images uploaded to SNS providers and
cloud photo storage services like google photos, due to manipulation on
providers, and to show that EtC systems are useful for such applications under
some conditions.  If we send encrypted images directly to receivers, this
difficulty will not be generated, but some advantages obtained by using the
providers, e.g. data storage services, will be lost. As a result, uploaded
images to such providers have currently no guarantee on privacy.

\begin{figure}[t]
\centering
\includegraphics[width =8.4cm]{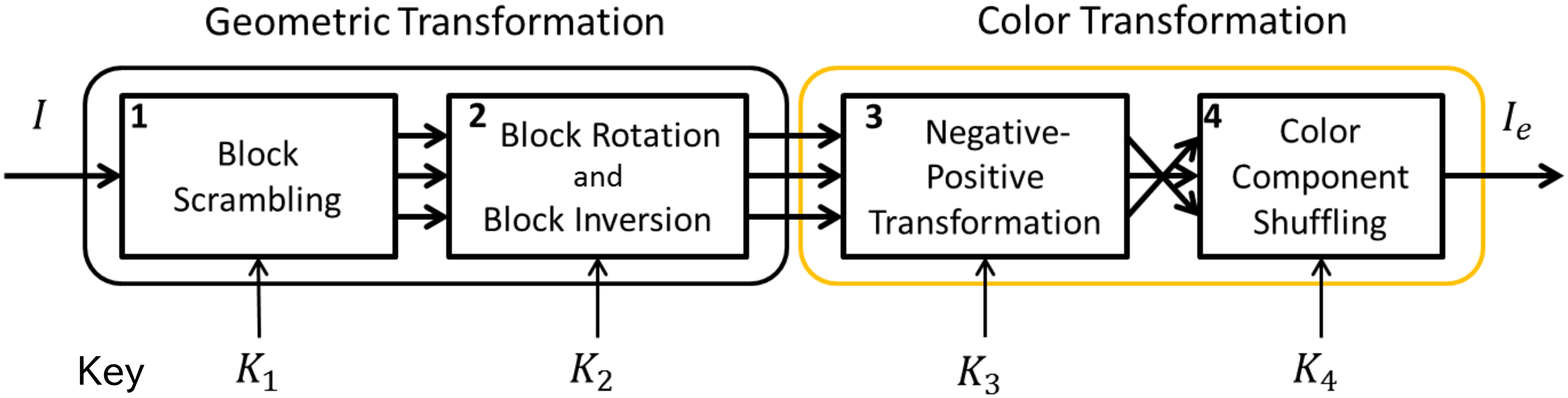}
\caption{Block scrambling-based image encryption}
\label{fig:step}
\end{figure}

\subsection{Block Scrambling-based Image Encryption}
Block scrambling-based image encryption schemes have been proposed for EtC
systems\cite{kurihara2015encryption,KURIHARA2015,KuriharaBMSB,Kuri_2017}. In
these
schemes\cite{watanabe2015encryption,kurihara2015encryption,KURIHARA2015,KuriharaBMSB,Kuri_2017},
an image with $X \times Y$ pixels is first divided into non-overlapping blocks with $B_x \times B_y$ pixels, with the number of blocks $n$ given by
\begin{equation}
n = \lfloor \frac{X}{B_x} \rfloor \times \lfloor \frac{Y}{B_y} \rfloor
\end{equation}
where $\lfloor \cdot \rfloor$ is the floor function that rounds down to the
nearest integer. Next, four block scrambling-based processing steps, as
illustrated in Fig.\,\ref{fig:step}, are applied to the divided image. The
procedure of performing image encryption to generate an encrypted image $I_e$
is as follows:

\begin{figure}[!t]
\centering
\hspace{2mm}\subfloat[Block rotation]{\includegraphics[width=3cm]{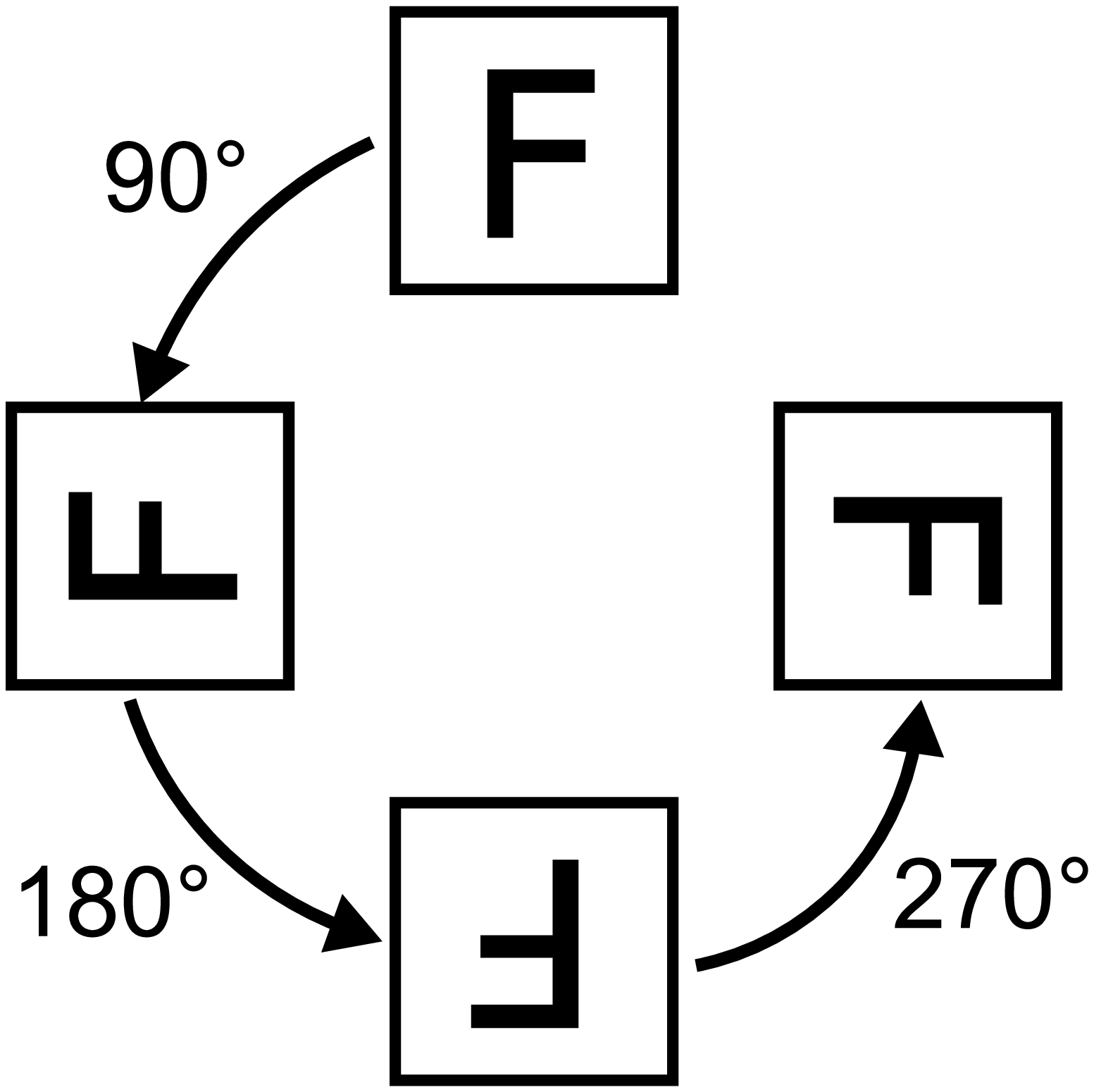}
\label{fig:label-B}}
\hfil
\subfloat[Block inversion]{\includegraphics[clip, width=3cm]{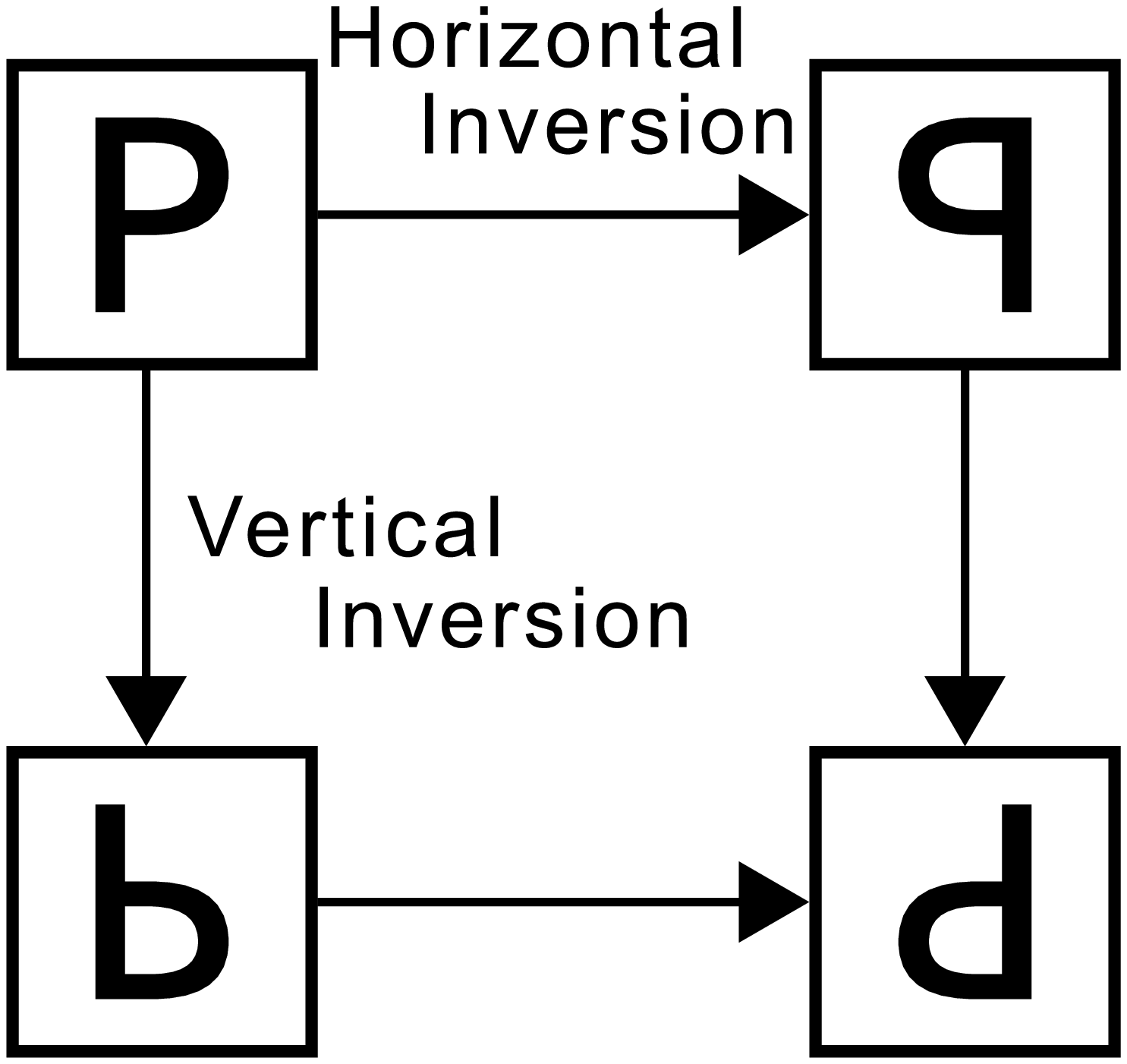}
\label{fig:label-C}}
\caption{Block rotation and inversion}
 \label{fig:rotinv}
\end{figure}

\begin{itemize}
\setlength{\parskip}{0cm}
\setlength{\itemsep}{0cm}
\item[Step 1:] Divide an image with $X \times Y$ pixels into blocks with $B_x
\times B_y$ pixels, and randomly permute the divided blocks using a random
integer generated by a secret key $K_1$, where $K_1$ is commonly used for all color components.
In this study, $B_{x}=B_{y}=16$. This is the same choice that was used in
\cite{KURIHARA2015}.
\item[Step 2:] Rotate and invert randomly each block (see
Fig.\,\ref{fig:rotinv}) by using a random integer generated by a key $K_2$,
where $K_2$ is commonly used for all color components as well.
\item[Step 3:] Apply the negative-positive transformation to each block by
using a random binary integer generated by a key $K_3$, where $K_3$ is commonly
used for all color components. In this step, the transformed pixel value in
the $i$th block $B_i$, $p'$, is computed as
\begin{equation}
p'=
\left\{
\begin{array}{ll}
p & (r(i)=0) \\
p \oplus (2^L-1) & (r(i)=1)
\end{array}
\right.
\end{equation}
where $r(i)$ is a random binary integer generated by $K_3$ and $p \in B_i$ is
the pixel value of the original image with $L$ bpp. \item[Step 4:] Shuffle
three color components in each block (color component shuffling) by using a
random senary integer generated by a key $K_4$.
\end{itemize}
\begin{comment}
\begin{figure}[!t]
\captionsetup[subfigure]{justification=centering}
\centering
\hspace{2mm}
\subfloat[Original image\cite{Nemoto_2014_MEX}\newline($X \times Y$ = $1920 \times 1080$)]
{\includegraphics[clip, width=3.9cm]{./image/19_ori.eps}
\label{fig:label-B}}
\hfil
\subfloat[Encrypted image\newline($B_{x}=B_{y}=16, n=8040$)]
{\includegraphics[clip, width=3.9cm]{./image/19_enc.eps}
\label{fig:label-C}}
 \\
 \hspace{2mm}
\subfloat[Decrypted image with block artifact (PSNR=31.4dB, sub-sampling ratio=4:2:0, downloaded from Facebook)]
{\includegraphics[clip, width=3.9cm]{./image/emm_noise2.eps}
\label{fig:label-B}}
\hfil
\subfloat[Decrypted image (PSNR=36.3dB, sub-sampling ratio=4:2:0, downloaded from Twitter)]
{\includegraphics[clip, width=3.9cm]{./image/emm_nonoise2.eps}
\label{fig:label-C}}
\caption{Examples of encrypted image and decrypted images, downloaded from Facebook and Twitter}
 \label{fig:oriencex}
\end{figure}
\end{comment}
\begin{figure}[!t]
\captionsetup[subfigure]{justification=centering}
\centering
\hspace{2mm}
\subfloat[Original image\cite{Nemoto_2014_MEX}\newline($X \times Y$ = $256
\times 144$)] {\includegraphics[clip, width=3.9cm]{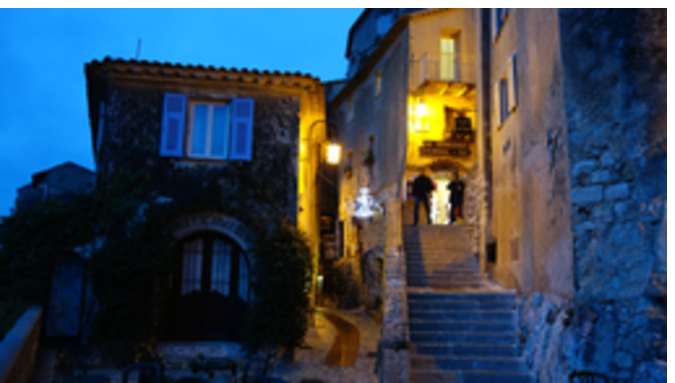}}
\hfil
\subfloat[Encrypted image\newline($B_{x}=B_{y}=16, n=144$)]
{\includegraphics[clip, width=3.9cm]{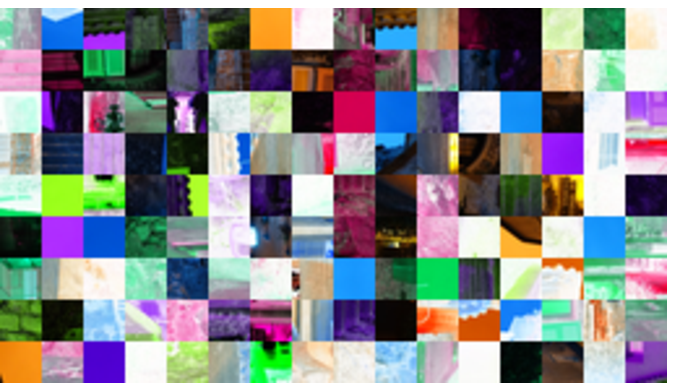}}
 \\
 \hspace{2mm}
\subfloat[Decrypted image with block artifact (PSNR=26.57dB, sub-sampling
ratio=4:2:0, downloaded from Facebook)] {\includegraphics[clip,
width=3.9cm]{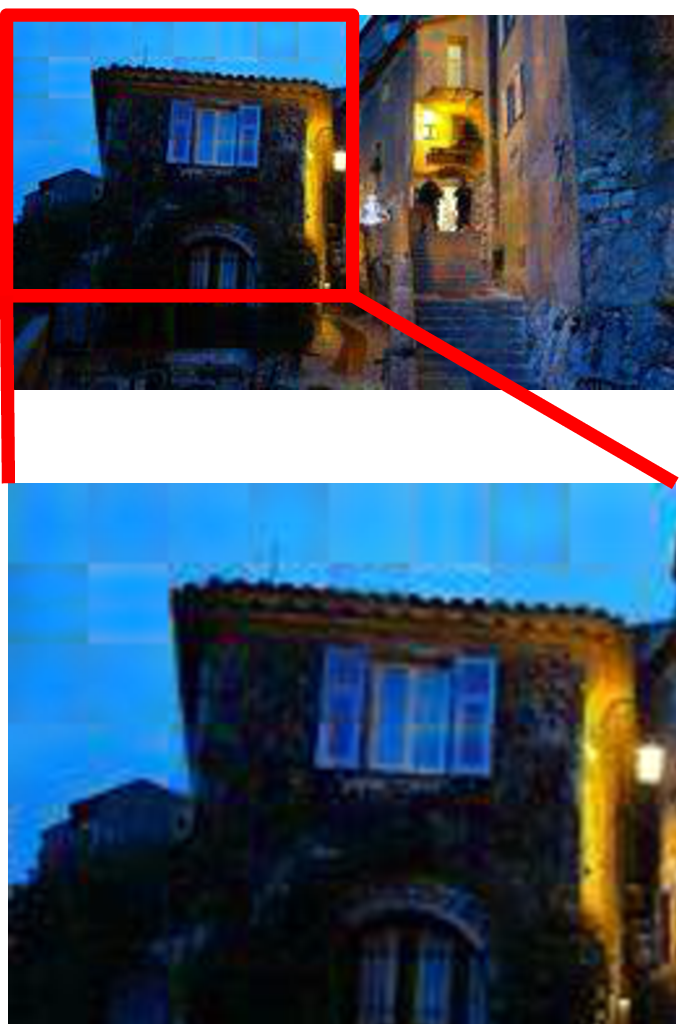}}
\hfil
\subfloat[Decrypted image (PSNR=29.37dB, sub-sampling ratio=4:2:0, downloaded
from Twitter)] {\includegraphics[clip, width=3.9cm]{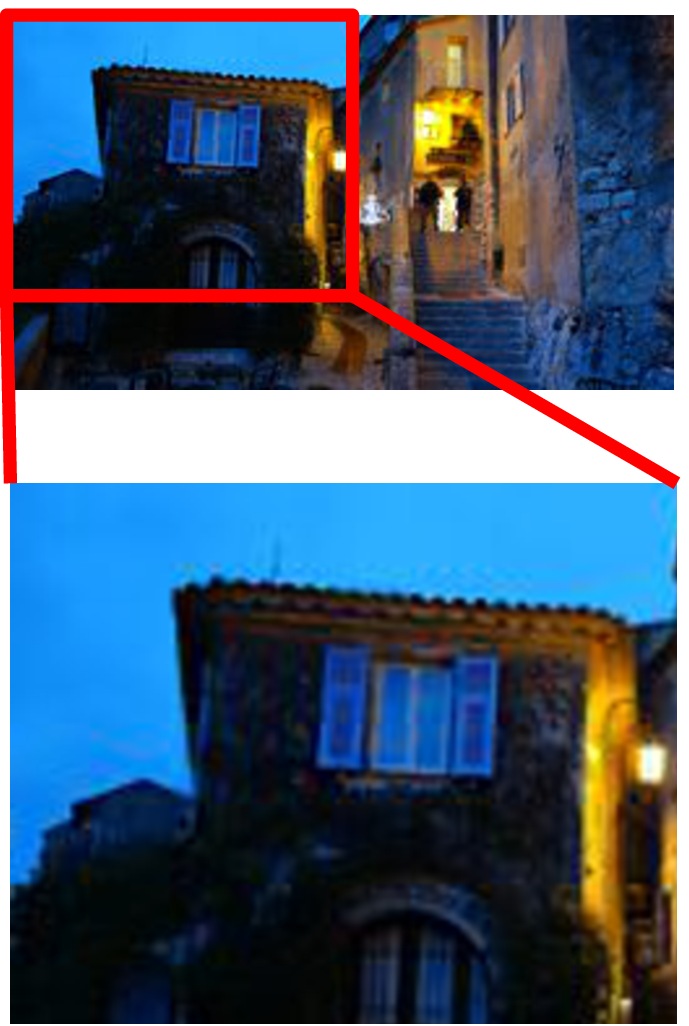}}
\caption{Examples of encrypted image and decrypted images, downloaded from Facebook and Twitter}
 \label{fig:oriencex}
\end{figure}

An example of an encrypted image is illustrated in
Fig.\,\ref{fig:oriencex}(b); Fig.\,\ref{fig:oriencex}(a) is the original
one. The key space of the block scrambling-based image encryption is generally
large enough to resist brute-force attacks\cite{KURIHARA2015}. On the other
hand, jigsaw puzzle solver attacks, which utilize correlations among pixels
in each block, have been
considered\cite{CHUMAN2017ICASSP,CHUMAN2017ICME,CHUMAN2017IEICE}. It is
confirmed that appropriate selection of the block size and combination
of encryption steps can improve the strength of EtC systems.

\begin{figure}[t]
\centering
\includegraphics[width =7.6cm]{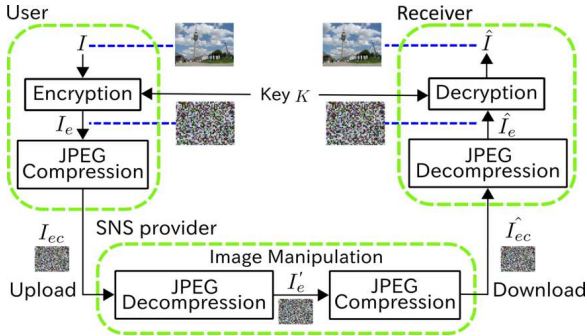}
\caption{EtC system}
\label{fig:etc}
\end{figure}

\subsection{Application to Social Media}
Fig.\,\ref{fig:etc} illustrates the scenario of this paper, where a user
wants to securely transmit an image $I$ to a receiver via an SNS provider.
Since the user does not give the secret key $K$ to the provider, the
privacy of the image to be shared is under the control of the user, even
when the provider decompresses it. Therefore, the user can ensure
privacy by him/herself. Even if the encrypted images saved on the SNS
servers are leaked by malicious users, third parties are not able to see them unless they have the key.
\par

Meanwhile, it is known that almost all SNS providers manipulate images uploaded
by their users, e.g., by rescaling the image resolution and
recompressing with different parameters, for decreasing their data size.
Manipulation of encrypted images by SNS providers might distort the
decrypted ones like in Fig.\,\ref{fig:oriencex}(c). Although numerous studies
have examined the conditions for resizing
images\cite{giudice2016_arxive,Moltisanti2015_ICIAP}, the actual
recompression parameters and conditions remain unpublished by SNS providers and
researchers. Therefore, we investigated how each SNS provider manipulates
images uploaded by users.

\section{Image Manipulation and Robustness}
In this section, we examine how each SNS provider manipulates images uploaded by
users. Then, the conditions to avoid block distortion are discussed with
regard to applying EtC systems to social media.

\subsection{Image Manipulation on Social Media}
We focus on two key aspects regarding image manipulation. The first aspect
is the maximum resolution of the uploaded images. The second is the
parameters of recompression.

\begin{figure*}[t]
\centering
\includegraphics[width =17cm]{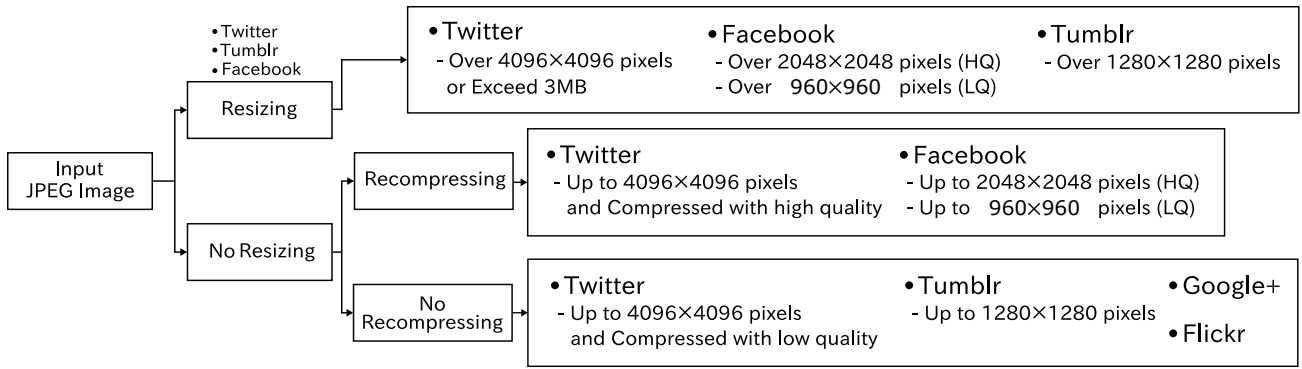}
\caption{Image manipulation on social media}
\label{fig:flowchart}
\end{figure*}

\subsection*{(1) Maximum Resolution}
A number of SNS providers automatically resize images uploaded by users when
the size of the images exceeds the maximum set by them
\cite{giudice2016_arxive,Moltisanti2015_ICIAP}. When the resolution is changed
in the encrypted domain, decrypting the images downloaded from providers becomes difficult.
\par
Fig.\,\ref{fig:flowchart} shows the classification of SNS providers in
terms of recompression and resizing. Twitter, Facebook, and Tumblr apply resizing
algorithms to uploaded images, if the images satisfy the following conditions.
%In contrast, Google+ and Flickr do not recompress uploaded images, and
%Tumblr also does not recompress uploaded images if the resolution is not
%exceed $1280 \times 1280$.
Twitter resizes uploaded images that are over 4096$\times$4096 pixels or exceed 3MB. Facebook has two modes to control the maximum resolution, i.e., Low Quality (LQ) and High Quality (HQ). The selection of LQ enables users to upload up to
images having 960$\times$960 pixels without any resizing. Meanwhile, HQ allows
them to upload images up to 2048$\times$2048 pixels. Tumblr changes the
resolution of uploaded images that are over 1280$\times$1280 pixels. Unlike
these three providers, Google+ and Flickr do not carry out any resizing
operations, even when the resolution of the uploaded images is large.

\subsection*{(2) Recompression}
Next, we investigated how each SNS provider recompresses images uploaded by
users. As illustrated in Fig.\,\ref{fig:oriencex}, the quality of downloaded
images depends on the provider. Block artifacts in the decrypted images might
be generated by recompression, as shown in Fig.\,\ref{fig:oriencex}(c).
\par

In terms of recompression, SNS providers are divided into two groups, as shown
in Fig.\,\ref{fig:flowchart}. Some providers, such as Google+, Tumblr, and
Flickr, manipulate only meta-data embedded in image files. Meanwhile,
Facebook recompresses all images regardless of the data size and
resolution.
Twitter recompresses images only when the data size of images uploaded by
users is larger than a threshold.
\par

Most SNS providers support JPEG \cite{JPEG_1991}, one of the most
widely used image compression standards. Therefore, we decided to focus on JPEG.
JPEG encoding of color images consists of six steps:
\begin{enumerate}
\item[1)]Perform the color transformation from RGB space to YCbCr space.
\item[2)]Sub-sample the Cb and Cr components to reduce the spatial resolution.
\item[3)]Divide the image into 8-by-8 blocks.
\item[4)]Apply the 2-D discrete cosine transform (DCT) to each block.
\item[5)]Carry out block-based quantizing with a quantization matrix \bm{$Q$}.
\item[6)]Carry out entropy coding using Huffman coding.
\end{enumerate}
SNS providers reduce the data size of uploaded images by changing the
quality factor $Q_{f}$(1$\leqq$$Q_{f}$$\leqq$100), which is a parameter to
control the matrix \bm{$Q$} in step 5). $Q_{f}$=100 gives the best quality, and $Q_{f}$=1 provides the worst quality.
\par

There are three sub-sampling ratios in the JPEG standard, referred to as 4:2:0
(reduction by a factor of 2 in both the horizontal and vertical directions),
4:2:2 (reduction by a factor of 2 in the horizontal direction), and 4:4:4
(no sub-sampling). The sub-sampling conditions are also important when
considering the effect of image manipulation by social media. The JPEG bitstream
of a color image is generated by performing steps 3) to 6) on the brightness
component Y and sub-sampled chroma components Cb and Cr independently.

\begin{table*}[t]
\centering
\caption{Relationship between uploaded JPEG files and downloaded ones in terms of sub-sampling ratios}
\begin{threeparttable}
\scalebox{1}{
\begin{tabular}{|c||c|c||c|c|}
\hline
\multirow{2}{*}{SNS provider} & \multicolumn{2}{c||}{Uploaded JPEG file} & \multicolumn{2}{c|}{Downloaded JPEG file} \\ \cline{2-5} 
 & \begin{tabular}[c]{@{}c@{}}Sub-sampling ratio\end{tabular} & $Q_{f}$ & \begin{tabular}[c]{@{}c@{}}Sub-sampling  ratio\end{tabular} & $Q_{f}$ \\ \hline
\multirow{4}{*}{Twitter (Up to 4096$\times$4096 pixels)} & \multirow{2}{*}{4:4:4} & low & \multicolumn{2}{c|}{No recompression} \\ \cline{3-5} 
 &  & high & 4:2:0 & 85 \\ \cline{2-5} 
 & \multirow{2}{*}{4:2:0} & 1,2,\ldots84 & \multicolumn{2}{c|}{No recompression} \\ \cline{3-5} 
 &  & 85,86,\ldots100 & 4:2:0 & 85 \\ \hline
\multirow{2}{*}{\begin{tabular}[c]{@{}c@{}}Facebook (HQ, Up to 2048$\times$2048 pixels)\\ 
Facebook (LQ, Up to 960$\times$960 pixels)\end{tabular}} & 4:4:4 & \multirow{6}{*}{1,2,\ldots100} & \multirow{2}{*}{4:2:0} & \multirow{2}{*}{71,72,\ldots85} \\ \cline{2-2}
 & 4:2:0 &  &  &  \\ \cline{1-2} \cline{4-5} 
\multirow{4}{*}{\begin{tabular}[c]{@{}c@{}}Tumblr (Up to 1280$\times$1280 pixels)\\ 
Google+\\ Flickr\end{tabular}} & \multirow{2}{*}{4:4:4} &  & \multicolumn{2}{c|}{\multirow{4}{*}{No recompression}} \\
 &  &  & \multicolumn{2}{c|}{} \\ \cline{2-2}
& \multirow{2}{*}{4:2:0} &  & \multicolumn{2}{c|}{} \\
 &  &  & \multicolumn{2}{c|}{} \\ \hline
\end{tabular}
}
\end{threeparttable}
\label{tb:change_sampling}
\end{table*}

\subsection{Recompression Parameters}
\label{recompress}
Although some image manipulation specifications on SNS providers have been
previously reported\cite{Caldelli_2017,giudice2016_arxive,Moltisanti2015_ICIAP},
the recompression conditions and recompression parameters are still unclear. In
the previous
works\cite{Caldelli_2017,giudice2016_arxive,Moltisanti2015_ICIAP}, the
maximum resolution condition to avoid resizing uploaded images, alternation of
image filenames and manipulation of JPEG headers have been investigated.
This paper investigates how the uploaded images are manipulated by the SNS providers in terms of the
color sub-sampling ratio and quality factor.

In the investigation, this paper focuses on 4:4:4 and 4:2:0 sub-sampling
ratios, although there are other sub-sampling ratios used in JPEG compression,
such as 4:4:0 and 4:2:2 sub-sampling ratios, because considering these two
ratios can determine the requirements to avoid block artifacts. Moreover, 4:4:4
and 4:2:0 sub-sampling ratios are widely used by most JPEG applications, including SNS
providers. Table\,\ref{tb:change_sampling} shows the relationship between
uploaded images and downloaded ones in terms of sub-sampling rate and quality factor.
This relationship, which has not been investigated before in any published
study, was confirmed by uploading and downloading a lot of JPEG images to
individual SNS providers through a personal computer. For instance, if a user
uploads JPEG images compressed with 4:4:4 sampling to Facebook, a receiver will view JPEG files manipulated with 4:2:0 sampling and certain
quality factors (71$\leqq$$Q_{f}$$\leqq$85). Note that $Q_f$ were estimated
by using JPEGsnoop software\cite{JPEGsnoop}, which utilizes the scaling
method from Independent JPEG Group (IJG)\cite{JPEGLIB} to obtain the
scaling factor used for generating the quantization table. Let us consider
image manipulation on Facebook and Twitter in more detail.

\subsection*{a) Image Manipulation on Facebook}
Facebook recompresses uploaded JPEG files with the sizes of up to
2048$\times$2048 (HQ) or up to 960$\times$960 (LQ) as follows.
\begin{enumerate}
\item[1)]Decompress JPEG files as color images in the spatial domain.
\item[2)]Compress the images at 4:2:0 sub-sampling ratio and specific
$Q_{f}$(71$\leqq$$Q_{f}$$\leqq$85) in accordance with the Facebook compression algorithm.
\item[3)]Save the recompressed JPEG files on a server to publish them for a receiver.
\end{enumerate}
In the above way, all uploaded images are converted to JPEG files with
4:2:0 sampling regardless of the data size of those images. As a result,
 the JPEG images with 4:2:0 color sub-sampling ratio are interpolated to
 increase the spatial resolution for chroma components in the decoding process.
 Since this interpolation process utilizes the relationship among blocks,
 encrypted images with 4:2:0 sub-sampling are affected by this interpolation.
 Hence, block artifacts are generated in the decrypted image, as shown in
 Fig.{\,\ref{fig:oriencex}(c)}.
\begin{table}[t]
\centering
\caption{Condition of JPEG images to prevent block artifacts from being
generated by Facebook and Twitter ($\circ$: Block artifact is generated,
$\times$: No block artifact)}
\label{fbtw}
\begin{tabular}{|c|cc|cc|}
\hline
SNS provider & \multicolumn{2}{c|}{Twitter} & \multicolumn{2}{c|}{Facebook} \\ \hline
\begin{tabular}[c]{@{}c@{}}Sub-sampling ratio\\
(Uploaded JPEG file)\end{tabular} & 4:4:4 & 4:2:0 & 4:4:4 & 4:2:0 \\ \hline
Block artifact & $\times$ & $\times$ & $\times$ & $\circ$ \\ \hline
\end{tabular}
\end{table}
\subsection*{b) Image Manipulation on Twitter}
Twitter recompresses uploaded JPEG files in accordance with the sub-sampling
ratio. When a user uploads JPEG files compressed at high quality and with
4:4:4 sampling to Twitter, the images are recompressed as follows.
\begin{enumerate}
\item[1)]Decompress JPEG files as color images in the spatial domain.
\item[2)]Compress the images at 4:2:0 sub-sampling ratio and $Q_{f}$=85.
\item[3)]Save the recompressed JPEG files on a server to publish them for a receiver.
\end{enumerate}
The image manipulation conditions of uploaded JPEG images with 4:4:4
sub-sampling depend on not only the uploaded quality factors, but also other
properties of images, so it is difficult to provide the strict definition of the
uploaded quality factor conditions. Thus, the condition of $Q_f$ is shown as
low/high in Table\,\ref{tb:change_sampling}. Twitter also recompresses
uploaded JPEG files if they were compressed under 4:2:0 sampling and high quality
($Q_{f}$$\geqq$85) as follows.
\begin{enumerate}
\item[1)]Reconstruct the DCT coefficients by using entropy decoding.
\item[2)]Quantize the DCT coefficients by using a quantization matrix
\bm{$Q$} with $Q_{f}$=85.
\item[3)]Carry out entropy coding using Huffman coding.
\item[4)]Save the recompressed JPEG files on a server to publish them for a receiver.
\end{enumerate}
Note that Twitter manipulates only meta-data, including in the header of
the uploaded JPEG images, if the images were compressed at a low
quality such as $Q_{f}$ = 60.

\subsection{Requirements to Avoid Distortion}
Decrypted images often have block distortion that depends on the
relationship between the encryption and recompression conditions. Here, we
examine how to avoid such distortion. In particular, we find that block
distortion does not result from image manipulation on social media if the
encrypted images satisfy the two conditions listed below.
\begin{enumerate}
\item[a)]The resolution of the encrypted images is left unchanged by the
SNS providers.
\item[b)]The encrypted images uploaded by users are compressed with 4:4:4 sub-sampling ratio.
\end{enumerate}
\par

Requirement a) means that the resolution of the encrypted images
needs to be smaller than the maximum resolution that each provider decides
as a resizing condition. Resizing the resolution of encrypted images makes
the block size of the encrypted images smaller, although the JPEG compression
is still carried out based on the size of $8 \times 8$. As a result each $16
\times 16$-block in resized encrypted images includes pixels from originally
different blocks, so the compression performance decreases and block distortion
is generated in the decrypted image due to the discontinuity among pixels.
Moreover, as shown in Fig.\,\ref{fig:flowchart}, users need not consider the
maximum resolution of encrypted images when uploading to Google+ and Flickr.
\par

Requirement b) means that we have to consider the sub-sampling ratios of
the encrypted images. Compression of JPEG images with 4:2:0
sub-sampling ratio is performed to increase the spatial resolution for chroma components in
the decoding process. This interpolation processing is carried out by using the
relationship among blocks. Therefore, the encrypted images compressed with
4:2:0 sub-sampling ratio are affected by this interpolation, while JPEG images
compressed with 4:4:4 sampling do not need any interpolation.
\par

However, JPEG files compressed with 4:2:0 sub-sampling ratio can sometimes
avoid block distortion even if interpolation is carried out.
Table\,\ref{fbtw} indicates the conditions under which JPEG files uploaded
by users will avoid block artifacts. As discussed in Sec.\,\ref{recompress},
Facebook performs interpolation in the spatial domain when JPEG images
compressed with 4:2:0 sampling ratio are uploaded by users. Consequently,
images with artifacts such as in Fig.\,\ref{fig:oriencex}(c) are
generated. Meanwhile, Twitter manipulates JPEG images in the DCT domain for
some operations such as quantization. Therefore, JPEG images that are
compressed with 4:2:0 sampling can avoid to block distortion due to
recompression when they are uploaded to Twitter. Thus, users do not need to
consider the sampling ratios of encrypted images when uploading to Twitter.

\section{Experimental Results}
We evaluated the effectiveness of EtC systems for social media by
conducting a number of simulations. In the simulations, encrypted and
compressed JPEG files were uploaded to SNS providers.

\subsection{Simulation Conditions}
The following procedure was carried out to evaluate the robustness of EtC systems based on Fig.\,\ref{fig:etc}.
\begin{enumerate}
\item[1)]Generate an encrypted image $I_{e}$ from an original image $I$ in accordance with Fig.\,\ref{fig:step}.
\item[2)]Compress the encrypted image $I_{e}$.
\item[3)]Upload the encrypted JPEG image $I_{ec}$ to SNS providers.
\item[4)]Download the recompressed JPEG image $\hat{I_{ec}}$ from the providers.
\item[5)]Decompress the encrypted JPEG image $\hat{I_{ec}}$.
\item[6)]Decrypt the manipulated image $\hat{I_{e}}$.
\item[7)]Compute the PSNR value between the original image $I$ and $\hat{I}$.
\end{enumerate}
\begin{figure}
\centering
\includegraphics[width =7.5cm]{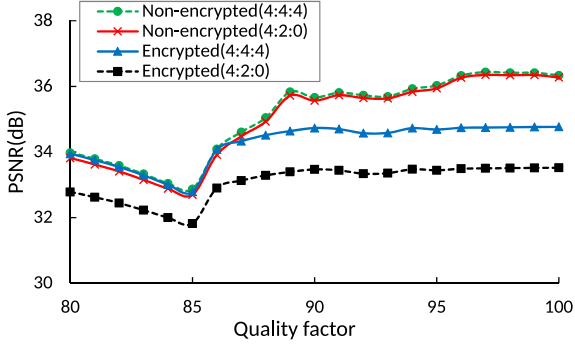}
\caption{Experimental result using original images as ground truth ones
(Facebook)}
\label{tb:fb_result}
\end{figure}
We made use of the JPEG implementation from IJG\cite{JPEGLIB} in steps 2) and 5). Then, we compressed each image with
4:2:0 or 4:4:4 sampling ratio and $Q_{f}=80,81\ldots100$. To compare the PSNR
values in step 7), the original image $I$ was compressed without any encryption
and then uploaded.
\begin{figure}
\centering
\includegraphics[width =7.5cm]{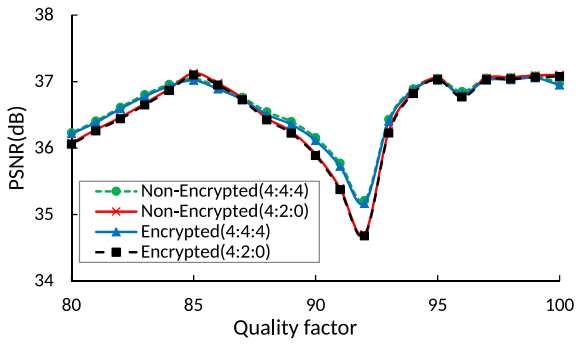}
\caption{Experimental result using original images as ground truth ones
(Twitter)}
\label{tb:tw_result}
\end{figure}
\par
To reduce dispersion, we used 20 FHD images from Ultra-Eye dataset
($1920 \times 1080$)\cite{Nemoto_2014_MEX}. We focused on Facebook and Twitter,
because these SNS providers recompress all images uploaded by users that meet
the conditions (see Fig.\,\ref{fig:flowchart}). The encrypted and non-encrypted
JPEG files compressed with 4:2:0 and 4:4:4 sampling were uploaded to the providers.

\subsection{Compression Performance of EtC System}
Figs.\,\ref{tb:fb_result} and \ref{tb:tw_result} show the experimental
results, where the average PSNR values of 20 images were calculated by
using original images without any compression distortion as ground truth ones.
The PSNRs of the decrypted images were low when images compressed with 4:2:0 sub-sampling
ratio were uploaded to Facebook. This shows that block distortion due to
recompression in the spatial domain greatly reduces the quality of decrypted
images. Even though the decrypted images with 4:4:4 sub-sampling ratio
did not include block distortion, their PSNRs were lower than
those of the non-encrypted images compressed with 4:4:4 sub-sampling
ratio, as indicated in Fig.\,\ref{tb:fb_result}. This is because Facebook
recompresses uploaded JPEG images with a non-constant
$Q_{f}(71,72,\ldots85)$, unlike Twitter.
\begin{figure}
\centering
\includegraphics[width =7.8cm]{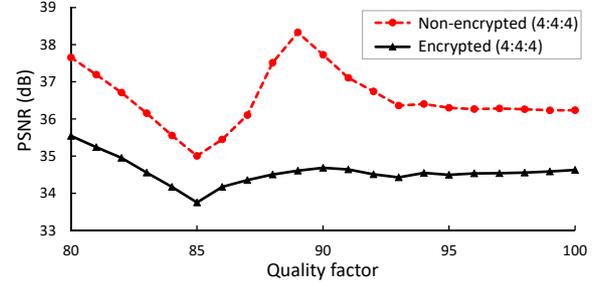}
\caption{Experimental result using JPEG encoded images as ground truth ones
(Facebook)}
\label{fig:single_as_ground}
\end{figure}
\begin{figure}[t]
\captionsetup[subfigure]{justification=centering}
\centering
\subfloat[Original image\cite{Nemoto_2014_MEX} ($X \times Y$ = $1920
\times 1080$)] {\includegraphics[clip, width=8cm]{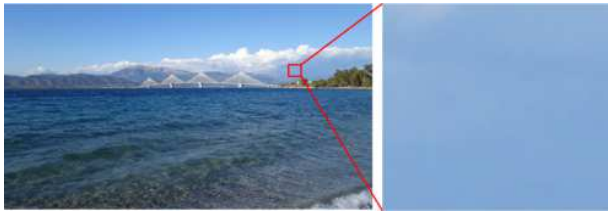}
}
\\
\subfloat[Decrypted image with block artifact (PSNR=29.89dB, downloaded from
Facebook, uploaded image with 4:2:0 sub-sampling and $Q_f=85$)]
{\includegraphics[clip, width=8cm]{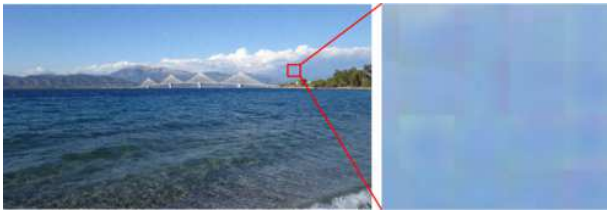} }
\\
\subfloat[Decrypted image (PSNR=35.99dB, downloaded
from Twitter, uploaded image with 4:2:0 sub-sampling and $Q_f=85$)]
{\includegraphics[clip, width=8cm]{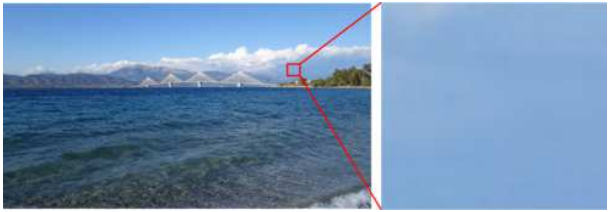} }
\caption{Examples of decrypted images, downloaded from
Facebook and Twitter}
\label{fig:best_worst}
\end{figure}
\par
On the other hand, Twitter recompresses images with 4:2:0 sub-sampling ratio in
the DCT domain. Thus, the PSNRs of the decrypted images uploaded to Twitter
were almost same as the non-encrypted ones even if the images were
compressed with 4:2:0 sub-sampling ratio. Regarding the decrypted images
with 4:4:4 sub-sampling ratio, the PSNRs were almost the same as the
non-encrypted ones.
\par
To clearly show the quality degradation caused
by the providers, in Fig.\,\ref{fig:single_as_ground}, PSNR values were
calculated by using JPEG encoded images as ground truth ones, which correspond
to images taken by a smartphone.
PSNRs of both non-encrypted and encrypted ones have almost the same tendency as
the result in Fig.\ref{tb:fb_result}.
\par
Moreover, the examples of an original image and
the decrypted images downloaded from Facebook and
Twitter are shown in Fig.\,\ref{fig:best_worst}. The result shows that the
decrypted image downloaded from Twitter did not include any block artifacts. In
contrast, block artifacts were generated in the decrypted image downloaded from
Facebook.

\section{Conclusion}
This paper proposed an application of EtC systems to enable users to
send images securely to receivers through SNS providers. Moreover, we
investigated how SNS providers manipulate JPEG images uploaded by users in
terms of their maximum resolution and recompression parameters.
%The results of a simulation showed that
In addition to the conditions that encrypted images uploaded by users generate some block distortion, the quality of images downloaded from the SNS providers was confirmed.
%some block distortion due to recompression by SNS
%providers greatly reduces image quality.
On the other hand, we determined that EtC systems are applicable to five SNS providers, Facebook, Twitter, Google+, Tumblr and Flickr, if the
encrypted images meet certain conditions.

\bibliographystyle{ieicetr}
\bibliography{refs}

\begin{thebibliography}{10}

\bibitem{huang2014survey}
C.T. Huang, L.~Huang, Z.~Qin, H.~Yuan, L.~Zhou, V.~Varadharajan, and C.C.J.
  Kuo, ``Survey on securing data storage in the cloud,'' APSIPA Transactions on
  Signal and Information Processing, vol.3, e7, 2014.

\bibitem{lagendijk2013encrypted}
R.~Lagendijk, Z.~Erkin, and M.~Barni, ``Encrypted signal processing for privacy
  protection: Conveying the utility of homomorphic encryption and multiparty
  computation,'' IEEE Signal Processing Magazine, vol.30, no.1, pp.82--105,
  2013.

\bibitem{zhou2014designing}
J.~Zhou, X.~Liu, O.C. Au, and Y.Y. Tang, ``Designing an efficient image
  encryption-then-compression system via prediction error clustering and random
  permutation,'' IEEE transactions on information forensics and security,
  vol.9, no.1, pp.39--50, 2014.

\bibitem{Zeng_2003}
W.~Zeng and S.~Lei, ``Efficient frequency domain selective scrambling of
  digital video,'' IEEE Transactions on Multimedia, vol.5, no.1, pp.118--129,
  2003.

\bibitem{Ito_2008}
I.~Ito and H.~Kiya, ``A new class of image registration for guaranteeing secure
  data management,'' IEEE International Conference on Image Processing (ICIP),
  pp.269--272, 2008.

\bibitem{Kiya_2008}
H.~Kiya and I.~Ito, ``Image matching between scrambled images for secure data
  management,'' 16th European Signal Processing Conference (EUSIPCO), pp.1--5,
  2008.

\bibitem{Ito_2009}
I.~Ito and H.~Kiya, ``One-time key based phase scrambling for phase-only
  correlation between visually protected images,'' EURASIP Journal on
  Information Security, vol.2009, no.841045, pp.1--11, 2010.

\bibitem{Tang_2014}
Z.~Tang, X.~Zhang, and W.~Lan, ``Efficient image encryption with block
  shuffling and chaotic map,'' Multimedia Tools Applications, vol.74, no.15,
  pp.5429--5448, 2015.

\bibitem{Erkin_2007}
Z.~Erkin, A.Piva, S.~Katzenbeisser, R.L. Lagendijk, J.~Shokrollahi, G.~Neven,
  and M.~Barni, ``Protection and retrieval of encrypted multimedia content:
  When cryptography meets signal processing,'' EURASIP Journal on Information
  Security, vol.2007, no.78943, pp.1--20, 2007.

\bibitem{nimbokar2014survey}
N.N. G. and S.S. V., ``Article: A survey based on designing an efficient image
  {Encryption-then-Compression} system,'' IJCA Proceedings on National Level
  Technical Conference X-PLORE 2014, vol.XPLORE2014, pp.6--8, 2014.

\bibitem{Johnson_2004}
M.~Johnson, P.~Ishwar, V.~Prabhakaran, D.~Schonberg, and K.~Ramchandran, ``On
  compressing encrypted data,'' IEEE Transactions on Signal Processing, vol.52,
  no.10, pp.2992--3006, 2004.

\bibitem{Liu_2010}
W.~Liu, W.~Zeng, L.~Dong, and Q.~Yao, ``Efficient compression of encrypted
  grayscale images,'' IEEE Transactions on Image Processing, vol.19, no.4,
  pp.1097--1102, 2010.

\bibitem{Zhang_2010}
X.~Zhang, ``Lossy compression and iterative reconstruction for encrypted
  image,'' IEEE Transactions on Information Forensics and Security, vol.6,
  no.1, pp.53--58, 2011.

\bibitem{Hu_2014}
R.~Hu, X.~Li, and B.~Yang, ``A new lossy compression scheme for encrypted
  gray-scale images,'' IEEE International Conference on Acoustics, Speech and
  Signal Processing (ICASSP), pp.7387--7390, 2014.

\bibitem{watanabe2015encryption}
O.~Watanabe, A.~Uchida, T.~Fukuhara, and H.~Kiya, ``An
  encryption-then-compression system for jpeg 2000 standard,'' IEEE
  International Conference on Acoustics, Speech and Signal Processing (ICASSP),
  pp.1226--1230, 2015.

\bibitem{kurihara2015encryption}
K.~Kurihara, S.~Shiota, and H.~Kiya, ``An encryption-then-compression system
  for jpeg standard,'' Picture Coding Symposium (PCS), pp.119--123, 2015.

\bibitem{KURIHARA2015}
K.~Kurihara, M.~Kikuchi, S.~Imaizumi, S.~Shiota, and H.~Kiya, ``An
  encryption-then-compression system for jpeg/motion jpeg standard,'' IEICE
  Transactions on Fundamentals of Electronics, Communications and Computer
  Sciences, vol.98, no.11, pp.2238--2245, 2015.

\bibitem{KuriharaBMSB}
K.~Kurihara, O.~Watanabe, and H.~Kiya, ``An encryption-then-compression system
  for jpeg xr standard,'' IEEE International Symposium on Broadband Multimedia
  Systems and Broadcasting (BMSB), pp.1--5, 2016.

\bibitem{Kuri_2017}
K.~Kurihara, S.~Imaizumi, S.~Shiota, and H.~Kiya, ``An
  encryption-then-compression system for lossless image compression
  standards,'' IEICE Transactions on Information and Systems, vol.E100-D, no.1,
  pp.52--56, 2017.

\bibitem{JPEG_1991}
G.~Wallace, ``The jpeg still picture compression standard,'' Communications of
  the ACM, pp.30--44, 1991.

\bibitem{Caldelli_2017}
R.~Caldelli, R.~Becarelli, and I.~Amerini, ``Image origin classification based
  on social network provenance,'' IEEE Transactions on Information Forensics
  and Security, vol.12, no.6, pp.1299--1308, 2017.

\bibitem{giudice2016_arxive}
O.~Giudice, A.~Paratore, M.~Moltisanti, and S.~Battiato, ``A classification
  engine for image ballistics of social data,'' arXiv preprint
  arXiv:1610.06347, 2016.

\bibitem{Moltisanti2015_ICIAP}
M.~Moltisanti, A.~Paratore, S.~Battiato, and L.~Saravo, ``Image manipulation on
  facebook for forensics evidence,'' International Conference on Image Analysis
  and Processing (ICIAP) 2015, pp.506--517, 2015.

\bibitem{Nemoto_2014_MEX}
H.~Nemoto, P.~Hanhart, P.~Korshunov, and T.~Ebrahimi, ``Ultra-eye: Uhd and hd
  images eye tracking dataset,'' Sixth International Workshop on Quality of
  Multimedia Experience (QoMEX), pp.39--40, 2014.

\bibitem{CHUMAN2017ICASSP}
T.~Chuman, K.~Kurihara, and H.~Kiya, ``On the security of block
  scrambling-based etc systems against jigsaw puzzle solver attacks,'' IEEE
  International Conference on Acoustics, Speech and Signal Processing (ICASSP),
  pp.2157--2161, 2017.

\bibitem{CHUMAN2017ICME}
T.~Chuman, K.~Kurihara, and H.~Kiya, ``Security evaluation for block
  scrambling-based etc systems against extended jigsaw puzzle solver attacks,''
  IEEE International Conference on Multimedia and Expo (ICME), pp.229--234,
  2017.

\bibitem{CHUMAN2017IEICE}
T.~Chuman, K.~Kurihara, and H.~Kiya, ``On the security of block
  scrambling-based etc systems against extended jigsaw puzzle solver attacks,''
  IEICE Transactions on Information and Systems, vol.E101-D, no.1, 2017.

\bibitem{JPEGsnoop}
C.~Hass, ``Jpegsnoop 1.8.0 - jpeg file decoding utility.''
  https://www.impulseadventure.com/photo/jpeg-snoop.html, 2017.

\bibitem{JPEGLIB}
``Independent jpeg group.'' http://www.ijg.org/.

\end{thebibliography}

\profile[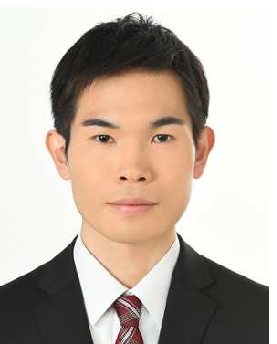]{Tatsuya Chuman}{
received his B.Eng. degree
from Toyo University, Japan in
2016. Since 2016, he has been a Master course
student at Tokyo Metropolitan University. His
research interests are in the area of image processing.}
\profile[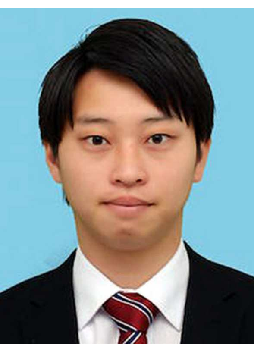]{Kenta Iida}{received his B.Eng. degrees from Tokyo Metropolitan University, Japan in 2016.
Since 2016, he has been a Master course
student at Tokyo Metropolitan University. His
research interests are in the area of image processing.}
\profile[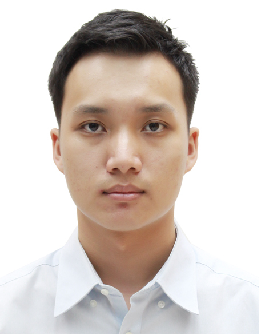]{Warit Sirichotedumrong}{received his B.Eng. and M.Eng. degrees from King Mongkut's University of Technology Thonburi, Thailand in 2014 and 2017, respectively.
Since 2017, he has been a Doctor course
student at Tokyo Metropolitan University. His
research interests are in the area of image processing.}
\profile[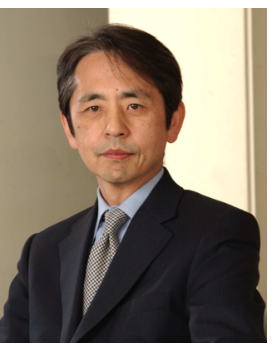]{Hitoshi Kiya}{received his B.Eng. and
M.Eng. degrees from Nagaoka University of Technology, Japan, in 1980 and 1982, respectively, and
his D.Eng. degree from Tokyo Metropolitan University in 1987. In 1982,
he joined Tokyo Metropolitan University as an Assistant Professor,
where he became a Full Professor in 2000. From 1995 to 1996, he
attended the University of Sydney, Australia as a Visiting Fellow. He
was/is the Chair of IEEE Signal Processing Society Japan Chapter, an
Associate Editor for IEEE Trans. Image Processing, IEEE Trans. Signal
Processing and IEEE Trans. Information Forensics and
Security. He also served as the President of IEICE
Engineering Sciences Society (ESS), the Editor-in-Chief for IEICE ESS
Publications, and a Vice President of APSIPA, He currently serves as
the President-Elect of APSIPA and Regional Director-at-Large for
Region 10 of the IEEE Signal Processing Society. He received the IEEE ISPACS
Best Paper Award in 2016, IWAIT Best Paper Award in 2014 and 2015, ITE
Niwa-Takayanagi Best Paper Award in 2012, the Telecommunications
Advancement Foundation Award in 2011, and IEICE Best Paper Award in
2008. He is a Fellow of IEEE, IEICE and ITE.}
\end{document}